# Role of nanophotonics in the birth of seismic megastructures

Stéphane Brûlé[1], Stefan Enoch[1] and Sébastien Guenneau[1]

[1] *Aix Marseille Univ, CNRS, Centrale Marseille, Institut Fresnel, Marseille, France*
*52 Avenue Escadrille Normandie Niemen, 13013 Marseille*
e-mail address: sebastien.guenneau@fresnel.fr

The discovery of photonic crystals thirty years ago, in conjunction with research advances in plasmonics and metamaterials, has inspired the concept of decameter scale metasurfaces, coined seismic metamaterials, for an enhanced control of surface (Love and Rayleigh) and bulk (shear and pressure) elastodynamic waves. The powerful mathematical tools of coordinate transforms, effective medium and Floquet-Bloch theories, which have revolutionized nanophotonics, can be translated in the language of civil engineering and geophysics. Experiments on seismic metamaterials made of buried elements in the soil demonstrate that the fore mentioned tools make possible a novel description of complex phenomena of soil-structure interaction during a seismic disturbance. But the concepts are already moving to more futuristic concepts and the same notions developed for structured soils are now used to examine the effects of buildings viewed as above surface resonators in megastructures such as metacities. But this perspective of the future should not make us forget the heritage of the ancient peoples. Indeed, we finally point out the striking similarity between an invisibility cloak design and the architecture of some ancient megastructures as antique Gallo-Roman theaters and amphitheatres.

## I. INTRODUCTION

One may wonder what is the link between seismic megastructures and Nanophotonics where photonics merges with nanoscience and nanotechnology, and where spatial confinement considerably modifies light propagation and light-matter interaction [1].

Seismic megastructures interact with decametric to hectometric wavelengths while nano-optics usually refers to situations involving ultraviolet, visible, and near-infrared light i.e. free-space wavelengths ranging from 300 to 1200 nanometers.

As a transposition of the definition of Nanophotonics, we would like to postulate that "seismic megastructures involve the science and engineering of mechanical wave-matter interactions that take place on wavelength and subwavelength scales where artificial structured matter controls the interactions". Two large scale experiments carried out in France in 2012 demonstrate it is possible to start the analysis of structured soils with the specific tools of condensed matter and, in particular, photonic crystals (PCs) and metamaterials [2, 3].

The study of complex wave phenomena in sub-surface soils, where many wave-matter interactions take place, is also progressing rapidly thanks to the development of three-dimensional surface sensors, which can be used in large numbers with good accuracy.

The spirit of data analysis has also changed a lot in the past few years. Indeed, the data processing in seismic prospection with artificial sources has been often governed by the oil and gas surface prospection approach, consisting in improving the signal to noise ratio, for a better tomography of the geological layers.

To identify the signal that contains the proofs of complex wave-matter interactions, it is necessary to modify the data acquisition and processing.

Among these geophysical techniques, we present here the on-site horizontal-to-vertical spectral ratio (HVSR) seismic method. HVSR is a non-invasive technique that can be used to estimate a dominant frequency of the soil covering the seismic substratum. Sedimentary deposits of the former Mexico lake are well-known to strongly amplify the seismic vibrations from the substratum to the free-surface of Earth. In this outstanding place, we distinguish the wind- and seismic-induced horizontal vibration of a 200 m-high tower and the low-frequency vibration of the soft soils by means of HVSR. This result corroborates the idea that the city could be studied under seismic disturbance as a set of locally resonant above-surface structures and put forward the recent concept of transformational urbanism and metacity.

In this review paper, we first present the main features of seismic megastructures made of elastic buildings (§2) and the soil-structure interaction (§3). Then we remind the main steps leading from electromagnetism to electrodynamics thanks to correspondences in the governing equations (§4), we extend the tool of transformation optics to seismology to design seismic cloaks (§5), we further consider the case of space folding transformations to cloak buildings outside seismic external cloaks (§6), we recall the principle of seismic ambient noise measurement (§7) and apply this geophysics technique to the site of Mexico city (§8), we discuss these results in light of similar phenomena arising in optics (§9) and finally draw some concluding remarks with some perspectives on the future of seismic metamaterials underpinned by research advances in nanophotonics (§10).

## II. SEISMIC MEGASTRUCTURES

Let us first stress that the seismic megastructures we describe in this article have metric to hectometric size. Basically, one can identify three main types of seismic metamaterials after a decade of research [4].

The first type includes structured soils made of cylindrical voids ([2] and [3]) or rigid inclusions ([5] and [6]), including seismic metamaterials. This group is called "Seismic Soil-Metamaterials" or SSM in short. The full-scale experiments with cylindrical holes described in [2] and [3] allowed the identification of the Bragg's effect similar to that in photonic crystals [2] and the distribution of energy inside the grid [3], which can be interpreted as the consequence of an effective negative refraction index (FIG. 1). The flat seismic lens in [3 **is** reminiscent of what Veselago and Pendry envisioned for light.

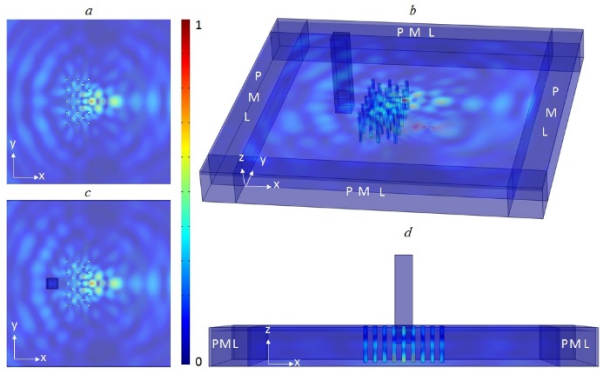

FIG.1. Finite element simulations for a thick plate with elastic (soil) parameters like those in figure 8 for a seismic source of time-harmonic frequency 12 Hz shows a shielding effect by the array of boheroles; (a) Top view of magnitude of displacement field for boreholes on their own. Note the vanishing displacement field amplitude just behind the array of boreholes. (b) 3D view for magnitude of displacement field for boreholes with a building (similar to that in figure 8) protected behind this array; (c) Top view of (b); (d) Side view of (b). Color scale ranges from dark blue (vanishing) to red (maximum) magnitude of displacement field and is normalized to the magnitude of the source.

The second group of seismic metamaterials consists of resonators buried in the soil in the spirit of tuned-mass dampers (TMD) like those placed atop of skyscrapers [7-10]. We call this group "Buried Mass-Resonators" (BMR). This type of structured soil modifies the initial seismic signal applied at the base of the surface structure. In Civil Engineering word, it acts on the kinematic effect, which, with the inertial effect, describes the soil-structure interaction [11].

The third type of seismic metamaterial consists of sets of Above-Surface Resonators (ASR), including any type of rigid elements clamped in the ground, from trees [12-14] to buildings [15]. It is specifically this type of structure that we present in this review paper (FIG 2).

Visionary research in the late 1980's based on the interaction of big cities with seismic signals and more recent studies on seismic metamaterials, made of holes or vertical inclusions in the soil, has generated interest in exploring the multiple interaction effects of seismic waves in the ground and the local resonances of both buried pillars and buildings.

Following the inspirational review article of Wegener's team at the Karlsruhe Institute for Technology [16], and ideas put forward by Housner [17, 18], Wirgin and Bard [19], Boutin [20], Philippe Guéguen [21] and his team at the Institute of Earth Science of Grenoble University observed that seismic noise could be modified in cities depending upon the specific arrangement and designs of tall buildings (FIG.2). Based on that idea, the concept of transformed metacities was launched in 2017 [15].
Interestingly, the idea of a dense urban habitat with high-rise buildings has already existed in the past. For instance, FIG. 2 is a realistic reconstitution of the medieval city of Bologna with its tall towers. Some of them are still in place today with a height of almost 100 m.

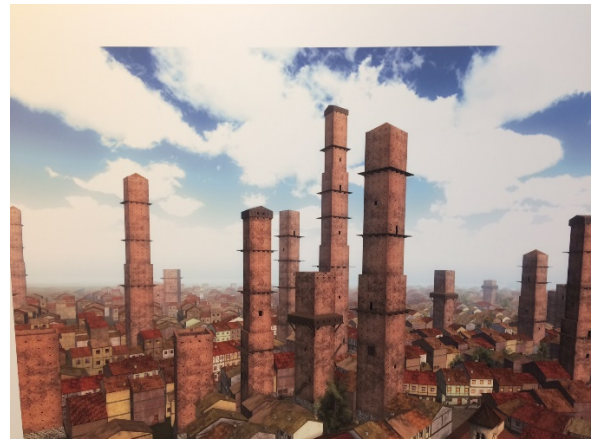

FIG. 2. Medieval towers of Bologna. Courtesy of The Time Machine by Genus Bononiae. Museum in the City (Bologna) and TOWER and POWER.

## III. MEGASTRUCTURES AND SOIL-STRUCTURE INTERACTION

The metacity is a concept first put forward in the late 1980's on the basis of experimental observations of ground response and its devastating effects. It has been known, since the 1950's, that the natural frequencies of any man-made structure are influenced by soil–structure interaction, especially on soft soils, and the presence of structures at the surface of a homogeneous half-space can significantly modify the ground motion [**22** to **25**]. Most of the time, the problem of ground response is disconnected from that of the resonant response of buildings or group of buildings.

On the basis of studies carried out on the interaction of cities with the seismic signal [**26**, **17**, **18**, **27**, **28**] and on the interaction of buildings with each other [**29**] (what could be viewed as a form of multiple scattering similar to that in PCs), some authors propose further extensions of this concept [**19**] based on analogies with electromagnetic and seismic metamaterials [**16**, **17**]. This theoretical approach is consistent with studies of several authors on the influence of the trees of a forest on the surface waves [**12** to **14**].

For the rest of our discussion we consider the building as an elastic element embedded in the ground. During the seismic disturbance, it remains in the elastic domain (FIG. 3).

Most of the vibration energy affecting nearby structures is carried by Rayleigh surface waves. Earthquake Engineering is concerned with the horizontal component of bulk and surface waves [**2**].

The response of a structure to an earthquake shaking is affected by interactions between three linked systems: the structure, the foundation, and the soil underlying and surrounding the foundation. Soil-structure interaction analysis evaluates the collective response of these systems to a specified ground motion. The terms Soil-Structure Interaction (SSI) and Soil-Foundation-Structure Interaction (SFSI) are both used to describe this effect ([**11**]).

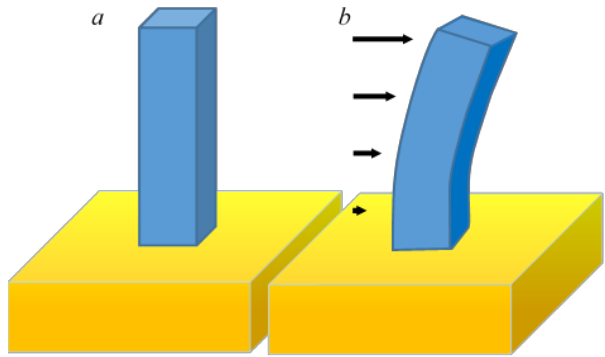

FIG. 3. Deformation of building on a thick plate: (a) Building at rest. (b) Elastic deformation for the first fundamental mode of the building.

The term free-field refers to motions that are not affected by structural vibrations or the scattering of waves at, and around, the foundation. SSI effects are absent for the theoretical condition of a rigid foundation supported on rigid soil.

We can summarize the objective of the design with the two schematics in FIG.4. Under the initial seismic disturbance the structure generates a significant inertial effect at the soil-structure interface (a). The objective with seismic metamaterials is to reduce the amplitude and/or modify the frequency of the vibrational excitation in order to reduce too the forces applied on the foundations and then, to improve their design (b). In both cases the inertial interaction could be described as a secondary seismic source emitted by the building itself. This is what we want to measure.

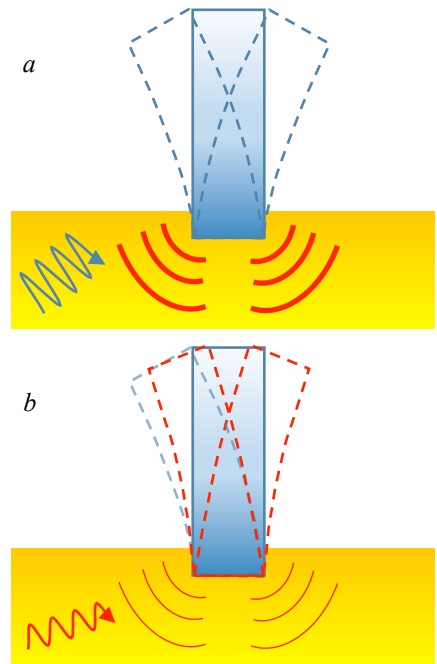

FIG. 4. Building under seismic disturbance and inertial interaction represented as a secondary seismic source, before (a) and after acting on the initial vibrational excitation (b). Acting on the input signal allows to influence, and particularly with the aim to decrease, the emitted vibration by the structure.

## IV. CORRESPONDENCES BETWEEN ELECTROMAGNETICS AND ELASTODYNAMICS

In 1987, the groups of E. Yablonovitch and S. John reported the discovery of stop band structures for light ([**30**] and [**31**]). Photonic crystals (PCs) have, since then, found numerous applications ranging from nearly perfect mirrors for incident waves whose frequencies fall in stop bands of the PCs, to high-q cavities for PCs with structural defects [**32**], and waveguides with line defects.

The occurrence of stop bands in PCs also leads to anomalous dispersion whereby dispersion curves can have a negative or vanishing group velocity. Dynamic artificial anisotropy, also known as all-angle-negative-refraction ([**33**] to [**36**]), allows for focusing effects through a finite

PC slab thanks to ray trajectories according to inverted Snell-Descartes laws of refraction, as envisioned in the 1968 paper of V. Veselago [37]. With the advent of electromagnetic metamaterials ([38] and [39]), J. Pendry pointed out that the image through the Veselago lens can be deeply subwavelength [40], and exciting effects such as simultaneously negative phase and group velocity of light [41], invisibility cloaks [42] and tailored radiation phase pattern in epsilon near zero metamaterials were demonstrated [43] and [44]. One of the attractions of platonic crystals, which are the elastic plate analogue of photonic and phononic crystals, is that much of their physics can be translated into platonics.

There are mathematical subtleties in the mathematical analysis, and numerical modelling, of the scattering of flexural waves [45] in elastic plates owing to the fourth-order partial derivatives present in the plate equations, versus the usual second-order partial derivatives for the wave equation of optics, involved in the governing equations; even waves propagating inside a perfect plate display marked differences compared to those of the wave equation as the former are not dispersionless, unlike the latter. Nonetheless, drawing parallels between platonics and photonics can be used as a guidance in the design of seismic metamaterials, as these bold parallels help achieve similar effects to those observed in electromagnetic metamaterials. Notably, dynamic effects reminiscent of the time dependent subwavelength resolution through a platonic flat lens [46], have been observed in the seismic flat lens [3], although further experimental evidence is required before arriving at hasty conclusions.

Actually, a research paper on phononic crystals provided numerical and experimental evidence of filtering effect for surface Rayleigh waves at MHz frequencies in a structured block of marble back in 1995 [47] and ten years ago subwavelength focusing properties [48] of acoustic waves via negative refraction have been achieved thanks to plasmon-like modes at the interfaces of a slab PC lens.

Localized resonant structures for elastic waves propagating within three-dimensional cubic arrays of thin coated spheres [49] and fluid filled Helmholtz resonators [50] paved the way towards acoustic analogues of electromagnetic metamaterials ([51] and [52]), including elastic cloaks ([53-55]). The control of elastic wave trajectories in thin plates was reported numerically [56] and experimentally in 2012 [2] and extended to the realm of surface seismic waves in civil engineering applications in 2014 [3].

Building upon analogies between the physics of flexural waves in structured plates and Rayleigh waves in structured soils to control surface seismic wave trajectories is not an incremental step. In order to achieve this goal, we had to solve conceptual and technological challenges. To name only a few, the duraluminium plate used in [2] is a homogeneous isotropic medium with simple geometric and elastic parameters, while the soil in [3] is heterogeneous and can only be ascribed some isotropic, homogeneous linear (e.g. non viscous) elastic parameters to certain extent, so its theoretical and numerical analysis requires some simplified assumption that can lead to inaccurate wave simulations. This means an experimental validation is absolutely necessary. Besides, Rayleigh waves are generated by anthropic sources such as an explosion or a tool impact or vibration (sledge-hammer, pile driving operations, vibrating machine footing, dynamic compaction, etc.), and this makes the numerical simulations even more challenging. Fortunately, an excellent agreement was noted between theory and experiment in [2], and this opens an unprecedented avenue in the design of large scale phononic crystals for the control of seismic waves, coined seismic metamaterials in [3].

In 1968, R.D. Woods [57] created in situ tests with a 200 to 350 Hz source to show the effectiveness of isolating circular or linear empty trenches, with the same geometry, these results were compared in 1988 with numerical modeling studies provided by P.K. Banerjee [58].

The main thrust of this article is to point out the possibility to create seismic metamaterials not only for high frequency anthropic sources, but also for the earthquakes' typical frequency range i.e. from 0.1 to 12 Hz.

Let us now go back to the roots of the seismic metamaterials' design, that counter intuitively rose from nanophotonics, with a typical frequency range in the Terahertz, so at the very least one trillion times ($10^{12}$) higher than for earthquakes.

## V. FROM TRANSFORMATION OPTICS TO TRANSFORMATION SEISMOLOGY

Actually, the field of transformational optics, which is 13 years old [59], can help explain metamaterial-like urbanism: when viewed from the sky, some modern cities and also older ones as in the medieval city of Bologna (FIG. 2), look similar to invisibility cloaks.

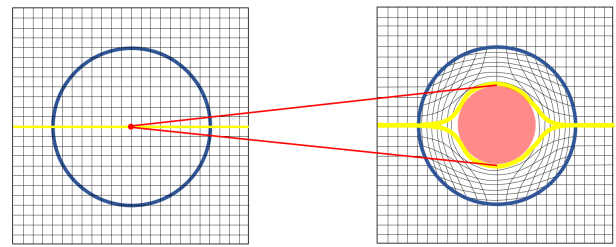

FIG. 5. Illustration of the transformation of a disk into a ring (inspired by [59]). The seismic ray in yellow passing through the center of the disk in the Cartesian coordinates (left) is detoured around the blown-up disk in the curvilinear coordinates (right). There is a direct analogy between ray trajectories in geometrical optics and seismology.

Let us consider a mapping from a disc of radius $r=R_1$ onto an annulus $R_1<r'<R_2$, known as Pendry's transform [59], as shown in FIG. 5

$$r' = R_1 + r\frac{R_2-R_1}{R_2} \qquad (1)$$

This actually stretches the coordinate grid from Cartesian (left panel) to curvilinear (right panel) inside the annulus. This local deformation creates some effective anisotropy in the transformed medium in the optics case [59], and besides from that some chirality in the elasticity case [54, 55]. Chirality can be achieved to certain extent by the local torsion of buildings placed atop a soil (buildings behaving like spinning elements of so-called micropolar elasticity theory). The required anisotropy can be achieved by a judicious placement of the buildings and group of buildings, as shown in FIG. 6.

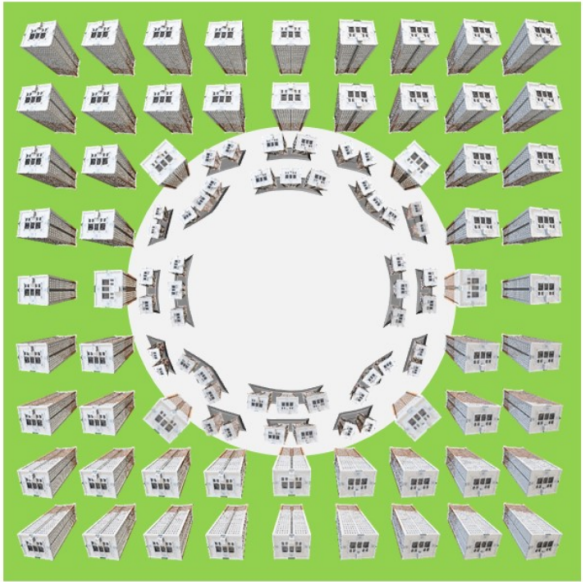

FIG. 6. Top view of a megastructure made of a set of above-surface bending resonators (cf. FIG. 4b). Here each resonator is a tall building. The typical size of the seismic cloak is 1 km.

We show in FIG. 7 an elastodynamic wave simulation (in-plane fully coupled pressure and shear waves in soil) performed with the commercial finite element package COMSOL MULTIPHYSICS. A source placed nearby a cloak with effective parameters similar to those in [54] i.e. with an anisotropic but also asymmetric elasticity tensor (the symmetry breaking being the hallmark of elastic chirality akin to magneto-optic activity), smoothly detour the pressure (left) and shear (right) wave polarizations around the disc in the center. However, one notes that the pressure component of the seismic wave does not vanish inside the disc. Fortunately, buildings are more resilient for pressure than shear vibrations, and the latter vanishes inside the disc, making it a safer zone to erect buildings. In terms of the megacity of FIG. 6, one can say that the buildings within the annulus of the cloak and outside the cloak would be badly affected by the seismic wave, but the white region in the center (e.g. a park) would be a safe zone where people could gather and remain safe during an earthquake. One could also envisage building a large structure (e.g. an amphitheater or a stadium) in the white area, and provided its dimensions are different from those of the buildings constituting the cloak, it would be protected from the shear component of the earthquake.

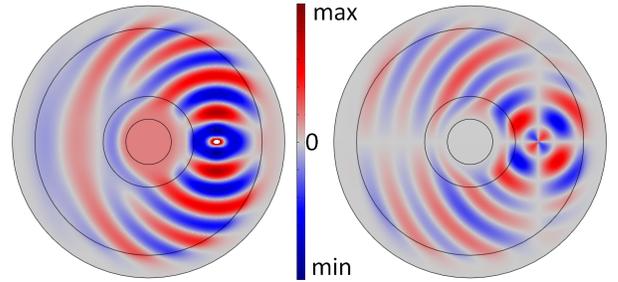

FIG. 7. Numerical simulation (COMSOL MULTIPHYSICS) for a seismic source placed in close proximity of a seismic cloak with effective elastic parameters compatible with the design in FIG. 6. Medium surrounding the cloak is soft soil, source frequency is 0.1 Hz and polarization of the source is along x (compressional wave). The horizontal (dilatational) displacement (left) is not completely damped inside the inner disc (soil), but the vertical (shear) displacement (right) indeed vanishes. Such a cloak therefore offers protection against shearing (most deleterious wave polarization) and any large structure (i.e. an amphitheater or a stadium) placed in the center region would be safe.

## VI. SPACE FOLDING FOR SCATTERING CANCELLATION OF CLAMPED INCLUSIONS

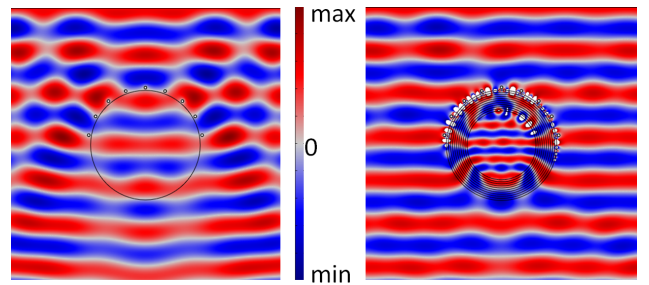

FIG. 8 Flexural plane wave propagating from top to bottom in a plate of thickness small compared with the wavelength. (Left) Scattering by 9 closely spaced clamped inclusion. (Right) Same with a seismic cloak consisting of

10 layers of isotropic homogeneous layers with a negative Young's modulus and a negative density deduced from the same space folding transform as in [60, 61]. One notes the significantly reduced scattering in the right panel.

As noted in FIG. 4b, a building can act as a secondary source, and thus can make an earthquake even more devastating as the combined effect of closely located building can create a strong seismic signal. One way to cancel the elastic signal of such antennae is to implement the concept of external cloaking, see FIG. 8., whereby negatively refracting isotropic elastic layers placed nearby 9 clamped inclusions (such as concrete columns in soil clamped to a bedrock creating a zero frequency stop band in [7]), reduce dramatically the scattering of a plane (flexural) incident wave. The same holds true for stress-free inclusions, such as boreholes in soil, or for buildings placed atop a soil. In fact, the flat seismic lens in FIG.1. can be viewed as an external cloak as well, since a negatively refracting index can always be associated with a space folding transform.

Let us now turn our mind to geophysics system, and the measurement techniques, in order to better bridge resonant phenomena for light in metamaterials with structured soil's vibrations.

## VII. PRINCIPLE OF SEISMIC AMBIENT NOISE HVSR

In this paragraph, the geophysical data are used only to show that the vibratory signature of some megastructures as skyscrapers could be identified in seismic noise acquisitions. We hope this can motivate further work.

**Principle of HVSR**

Geological site conditions can generate significant changes in earthquake ground motion producing concentrated damage ([62] and [63]). It is usual to characterize the site effects by means of spectral ratios of recorded motions with respect to reference rock site [64]. These ratios are called empirical transfer functions can be easily obtained for seismically active locations. It is often appropriate to interpret this transfer function assuming a 1-D soil configuration and deal with resonant frequencies and amplifications (FIG. 9). Empirical transfer function could be compared to theoretical curves (FIG. 10).

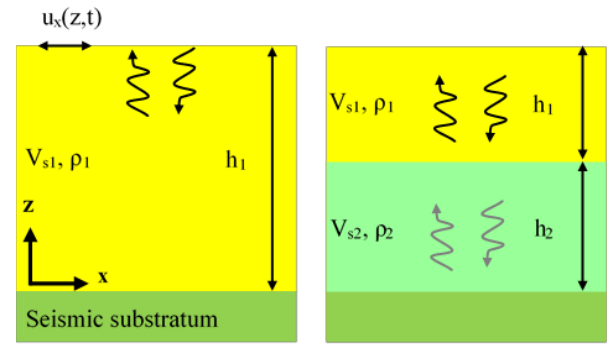

FIG. 9. Schematics of wave propagation in a 1D soil-model. Homogeneous (left), two-layers model (right).

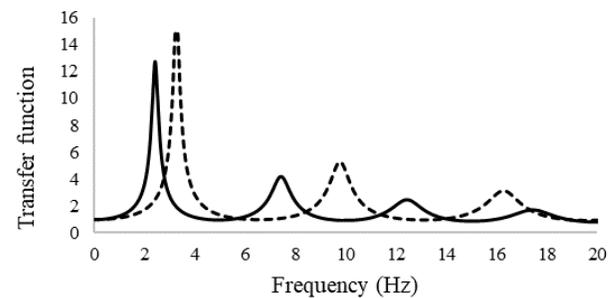

FIG. 10. 1D soil-model. Transfer function for the theoretical displacement at the seismic substratum and the free surface. The thickness of soil deposits is 20 m. The shear wave velocity $V_s$ is 200 m.s$^{-1}$ (black solid line) and 260 m.s$^{-1}$ (black dotted line). The damping ratio is respectively 0.05 and 0.04.

Initiated by Nakamura in 1989 [65] and widely studied to explain its strengths and limitations (seismological SESAME research project, [66], [67], [68], [69] and [70]), the seismological microtremor horizontal-to-vertical spectral ratio (HVSR) method offers a reliable estimation of the soil resonance frequencies from the spectral ratio between horizontal and vertical motions of microtremors. This technique usually reveals the site dominant frequency $f_0$ but the amplitude of the HVSR is still not well understood [36].

The HVSR method was tested in geotechnical engineering too ([70], [71] and [72]) by virtue of its ease of use. In restrictive conditions, as for homogeneous soil layer model with a sharp acoustic impedance contrast with the seismic substratum, the technique could be employed to estimate the ground fundamental frequency $f_0$. In this paper, we use this technique to point out both the fundamental low frequency of the alluvial basin and the frequencies of the LatinoAmericana Tower in Mexico City.

# VIII. DESCRIPTION OF THE SITE AND THE HVSR MEASUREMENT SURVEY

**Geology of Mexico**

The subsoil of the Mexico Valley is known worldwide for its lacustrine clays (high water content) and its seismic site effects, especially after the dramatic major earthquake of September 19$^{th}$, 1985 which led to more detailed studies [73]. Typically, the geotechnical zonation of Mexico City is made up of three areas: hills, transition and ancient lake zone. The LatinoAmericana Tower is located on the lake bed zone where the thickness of clay exceeds 30 m (FIG.11 and FIG. 12). At the western part of the valley, the thickness of lacustrine deposits could reach 60 m. Shear wave velocity Vs for the clays typically presents very low values (lower than 100 m/s).

**Torre LatinoAmericana**

Inaugurated in 1956, The LatinoAmericana Tower (FIG. 11.) is a 182 m high skyscraper with 44$^{th}$ floors. The square base is 34.4 x 34.4 m and the overall weigth of the tower is around 23 500 tons.

Its design consists of a steel-frame construction (H columns) and 361 deep-seated piles anchored at -33 m below the street level, in a little thick sand layer (5 m).

Considered as the most resistant (or one of the most resistant) building of Mexico City, it survived three destructive earthquakes: 1957/07/28 (moment magnitude M$_w$: 7.7), 1985/09/19 (moment magnitude M$_w$: 8.0) and 2017/09/19 (Magnitude M$_w$: 7.1). We can thus confer to this structure the merit for the tallest building ever exposed to huge seismic forces.

The ground surface subsidence problem of Mexico City necessitated a 13 meters deep basement to reduce the net bearing pressure on the piles raft. The foundation is a rigid mat supported by piles ([74] and [75]). The estimation of the different modes of vibration of the building gives 3.5 to 3.7 s for the fundamental, 1.5, 0.9 and 0.7 s for the harmonics [74]. The ground period for an earthquake is estimated between 1.5 to 2.5 s in the downtown which is located in the ancient lake area. Basically, an excitation of the second mode of the building is expected during an earthquake.

**HVSR Measures**

We chose two special locations for the HVSR measures with the aim to have the same soil fundamental frequency f$_0$ on both graphs and to point out the characteristic frequency of the tower vibration. For that, the reference measure was recorded in the central part of a plant park with no buildings within 200 m of distance, see FIG. 13. The measure for the building was taken outside at 1 m from the building wall of the ground floor, on a concrete slab., as indicated by the arrow on FIG. 11. Each measure was double-recorded with two independent sensors. The horizontal distance between the two spots is 415 m.

The present study used a highly portable (10 x 14 x 7.7 cm (height), 1.1 kg) three component (2 horizontal and 1 vertical directions) high-resolution electro-dynamic sensor (Tromino$^{TM}$ from Micromed) for the measurement of the ambient vibrations at the observation point or station. The technical characteristics of the sensor are given in Table 1.

The HVSR measurements in this study were made in compliance with the guidelines of SESAME [76], so the step-by-step methodology will not be repeated here except to highlight a few salient points. Before taking the measurement, the sensor was aligned and firmly secured to the ground by a set of spikes. Seismic noise was recorded with a sampling rate of 512 Hz for 30 min at each site, this is to ensure that there is adequate statistical sampling in the range (0.1–100) Hz, the frequency range of engineering interest.

The HVSR curves were calculated by averaging the H/V (horizontal-vertical) ratios obtained after dividing the signal into non-overlapping 20-s windows (which is sufficient for the spectra above 1 Hz). Each window was detrended, tapered, padded, FF-transformed and smoothed with triangular windows of width equal to 10% of the central frequency.

The ratio of the H/V Fourier amplitude spectra is expressed with (Eq.2), where $F_V(\omega)$ and $F_H(\omega)$ denote the vertical and horizontal Fourier amplitude spectra respectively.

$$HVSR(\omega) = \frac{|F_H(\omega)|}{|F_V(\omega)|} \qquad (2)$$

The geometric average was used to combine EW and NS components in the single horizontal(H) spectrum, and then it was divided by the vertical component (V) to produce the measured HVSR curve as shown in (Eq. 3).

$$HVSR_m = \frac{\sqrt{H_{EW} \times H_{NS}}}{V} \qquad (3)$$

Furthermore, the stability of the HVSR curves was verified and used to identify the presence of artifacts from anthropic noise, (machinery, industrial areas, etc.).

| Velocimeter specifications | |
|---|---|
| Frequency response | 0.1 – 256 Hz |
| Dynamic range | 180 Db |
| Sample rate per channel | 32 kHz |
| Output sampling rate | 128, 256 or 512 Hz |

TAB. 1. Digital velocimeter specifications.

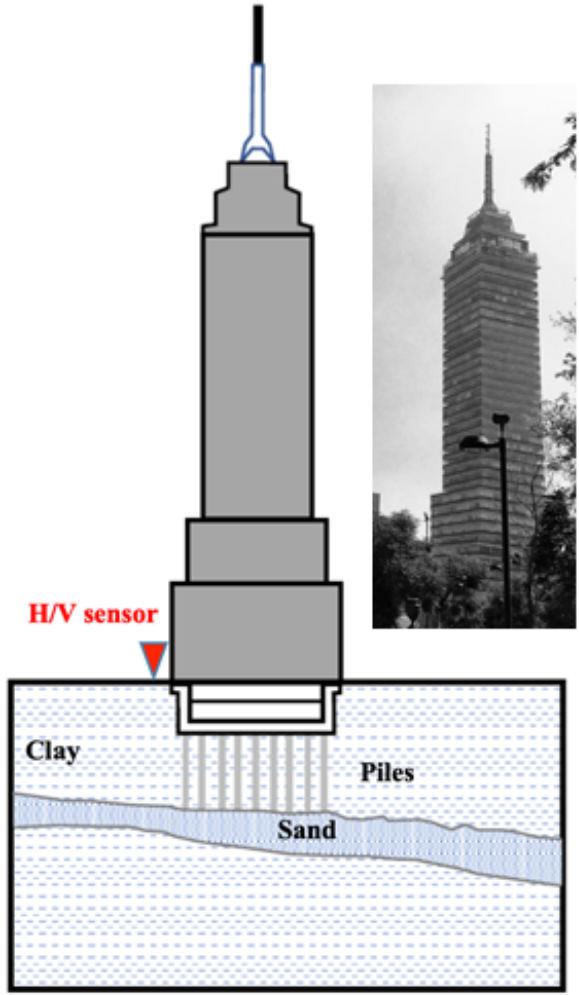

FIG. 11. Schematic of Torre LatinoAmericana (182 m) and its deep pile-foundations. The soil consists of clay, with a layer of sand, with an overall thickness over 30 m. The location of the velocimeter, whose specifications are given in Table 1, is indicated by the arrow. Photography courtesy of S. Brûlé.

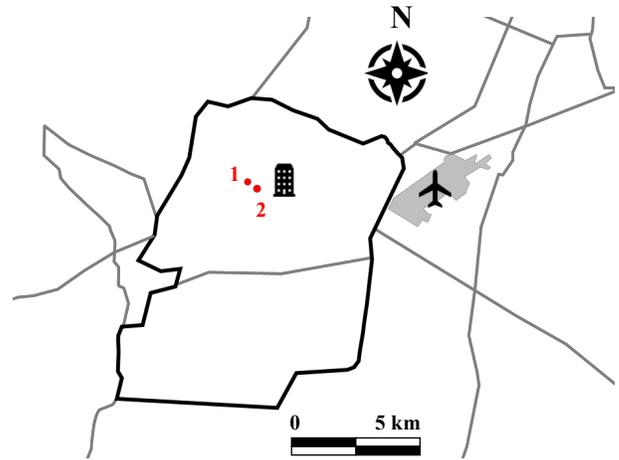

FIG. 12. Schematic map of Mexico City Center and location of HVSR measures (1: reference measure; 2: tower measure).

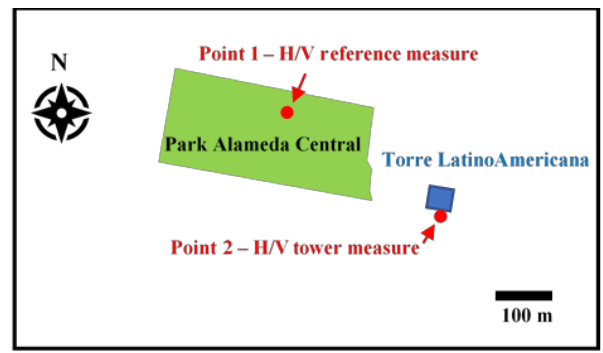

FIG. 13. Location of HVSR measures in Mexico City. The horizontal distance separating 1 and 2 is 415 m.

## IX. RESULTS AND DISCUSSIONS

Results are presented in FIG. 14, TAB. 2 and TAB. 3.

In the frequency range of engineering interest, it can be noted first that a clear distinctive peak appears for all the measures and the amplification is 5 times, at least.

Fine lines curves represent the 95% of interval of confidence. These curves are very close to the average solid red or green curve, indicating a good quality of measurement.

For the point 1 in the park, the average period (sensors 70 and 71) at the dominant frequency is 1.52 s. For the point 2, close to the tower, this average period is 1.42 s. We remind that a horizontal distance of 415 m separates the two points.

The overall geology of the basin is known to be stable between these two points, but local variations may exist.

For the measure 2, a second frequency peak at 0.45 Hz (sensor 70) and 0.42 Hz (sensor 71) is observable on FIG. 14. The mean period is 2.3 s. This period around 2 s could be the signature of the vibration generated by the tower itself on its close environment. We do not distinguish the fundamental period of the tower (3.5 to 3.7 s) nor its harmonics (1.5, 0.9 and 0.7 s). However, it should also be kept in mind that the many earthquakes experienced by a structure may have changed the values of the initial modes as well as the ground-foundation-structure interface conditions.

Frequency peaks at frequencies above 10 Hz could be attributed to forest of trees in the park, as these present similarity with filtering effects observed and numerically justified for a forest of trees in Grenoble [**12**].

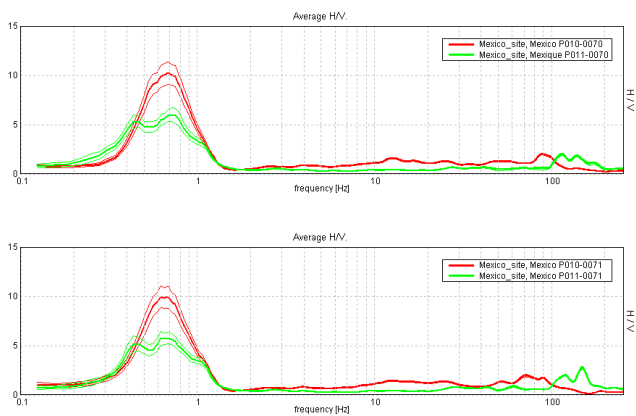

FIG. 14. HVSR Results. Measure 1 and 2 in same graph for sensor 70 (Top) and for sensor 71 (Bottom)

| Spot | Latitude | Longitude | City | Location |
|---|---|---|---|---|
| 1 | 19,4361 | -99,1436 | Mexico | Alameda Central |
| 2 | 19,4337 | -99,1406 | Mexico | Torre LatinoAmericana |

TAB. 2. Location of the HVSR measures.

| Spot | Sensor | First dominant frequency | T at dominant frequency | Second dominant frequency | T at dominant frequency |
|---|---|---|---|---|---|
| | | Hz | s | Hz | s |
| 1 | 70 | 0,69 +/- 0,11 | 1,45 | | |
| 1 | 71 | 0,63 +/- 0,11 | 1,59 | | |
| 2 | 70 | 0,72 +/- 0,11 | 1,39 | ~0,45 | 2,22 |
| 2 | 71 | 0,69 +/- 0,11 | 1,45 | ~0,42 | 2,38 |

TAB. 3. HVSR results.

## X. CONCLUSIONS AND PERSPECTIVES

Our voyage in the wonderworld of seismic metamaterials nears the end. We have seen that it is rooted in research advances in nanophotonics. The radical change of scale from nanometers to decameters suggests that many analogies drawn between light and seismic wave phenomena break down, but large scale experiments come to our rescue and backup the bold claim of negative refraction, lensing and cloaking taking place in soft soils structured at the decameter scale. While working on this review article, we realized that there are striking similarities between an invisibility cloak tested for various types of waves [**77**] and sky views of antique Gallo-Roman theaters [**78**, **79**, **80**], see Fig. 15.

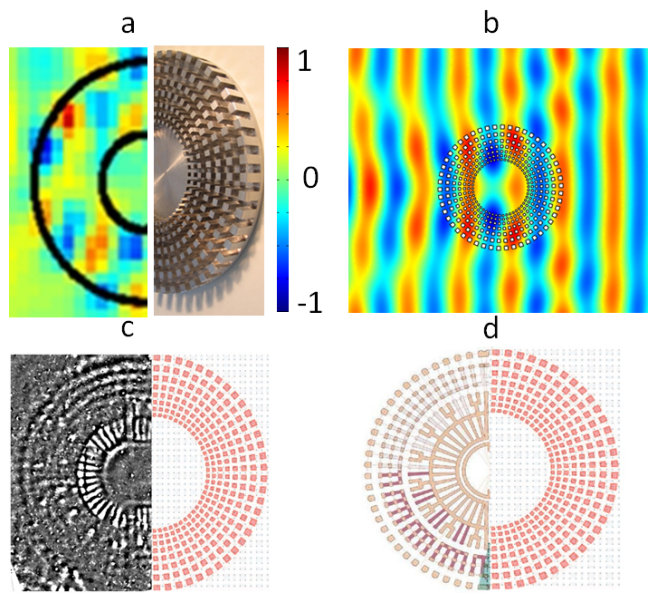

FIG. 15. Analogy between an invisibility cloak and antique Gallo-Roman theaters: (a) Electric field from microwave experiment at 3.5 GHz (courtesy of R. Abdeddaim and E. Georget) and photo (CNRS/INSIS/Institut Fresnel) of an aluminium cloak 20 cm in diameter. (b) Numerical simulation (COMSOL MULTIPHYSICS) of a plane flexural wave incident from the left at 5 Hz on multiply perforated cloak [**56**] 200 m in diameter in a 5m thick plate consisting of clay. (c) Photo montage of a sky geophysics view extracted from a magnetic gradient map of a completely buried Gallo-Roman ex muros theater located at Autun, La Genetoye (France) versus the cloak (courtesy of Geocarta and G. Bossuet, in [**78**] with the permission of the authors). (d) Same comparison with a part of the foundations of the antique theater (courtesy of A. Louis (PCR) and Hydrogéotechnique [**79**]).

It seems to us that such architectures are more resilient thanks to their unique design. Perhaps this is the reason why some of these megastructures, as amphitheaters, have remained mostly intact through the centuries? This deserves to be studied in more detail, and we hope our preliminary observations will foster efforts in the search for more analogies between engineered electromagnetic and and a new type of seismic metamaterials we would like to call 'archeological metamaterials'.

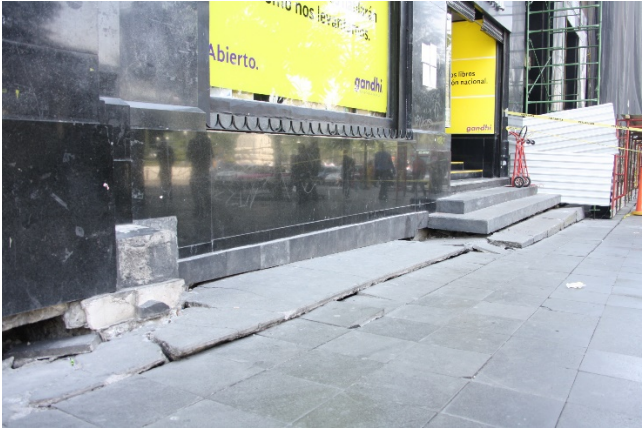

FIG. 16. Evidence of soil-structure interaction in Mexico City (Photo courtesy of S. Brûlé and A.F.P.S. [81]).

On the contrary, recent constructions such as the one shown in FIG. 16. do not share the same resilient features as Roman amphitheaters. The future of seismic metamaterials might be brighter if physicists and civil engineers, not only draw useful analogies with design of electromagnetic metamaterials and metasurfaces, but also take a closer look at ancient architectures and learn from these beautiful and amazingly resilient designs.

## ACKNOWLEDGEMENTS

S.G. would like to acknowledge the group of Prof. R.V. Craster at Imperial College London for a visiting position. S.B. thanks especially Association Française de Génie Parasismique (A.F.P.S.) [81] that has organized the Mexico mission in 2017, C. Boutin, S. Hans and T. Doanh from Ecole Nationale des Travaux Publics de l'Etat at Vaulx en Velin, France, for having loaned him H/V sensors. For archeological material and permission for figures, S.B. would like to acknowledge all the authors involved in PCR Project [80], and more personally, Y. Labaune (Director of "Service Archéologique de la Ville d'Autun, France"), A. Louis (Direction de l'Aménagement, Service de l'Archéologie, Conseil Départemental d'Eure-et-Loir, Chartres, France), J-B and J-C Gress (Direction of Hydrogéotechnique) H. Grisey (Technical Direction at Hydrogéotechnique), L. Colin (Université de Franche-Comté and Director of the subsidiary of Hydrogéotechnique at Belfort) and F. Ferreira (Director of archeological excavation at La Genetoye, Autun).